\documentclass[11pt,a4paper]{article}

\usepackage[utf8]{inputenc}
\usepackage{authblk}
\usepackage[numbers]{natbib}
\usepackage{hyperref}
\usepackage{graphicx}
\usepackage{booktabs}
\usepackage{amsmath}
\usepackage{makecell}
\usepackage{adjustbox}

\newcommand{\orcid}[1]{\href{https://orcid.org/#1}{#1}}

\title{Rural Connectivity Inequalities in Finland and Sweden: Evidence, Measures, and Policy Reflections}

\author[1]{Sameera Bandaranayake\thanks{\orcid{0009-0005-9776-0827}}}
\author[2]{Amirreza Moradi\thanks{\orcid{0009-0009-0332-3220}}}
\author[4]{Tanja Suomalainen\thanks{\orcid{0000-0001-8966-3932}}}
\author[1]{Harri Saarnisaari\thanks{\orcid{0000-0001-9857-3234}}}
\author[1,3]{Pasi Karppinen\thanks{\orcid{0000-0002-7800-8243}}}
\author[2]{Payal Gupta\thanks{\orcid{0000-0003-0726-065X}}}
\author[2]{Jaap van de Beek\thanks{\orcid{0000-0001-8647-436X}}}

\affil[1]{University of Oulu, Finland}
\affil[2]{Luleå University of Technology, Sweden}
\affil[3]{Vrije Universiteit Brussel, Belgium}
\affil[4]{Lapland University of Applied Sciences, Finland}

\date{}

\begin{document}
\maketitle

\begin{abstract}
Persistent rural–urban disparities in broadband connectivity remain a major policy challenge, even in digitally advanced countries. This paper examines how these inequalities manifest in northern Finland and Sweden, where sparse populations, long distances, and seasonal variations in demand create persistent gaps in service quality and reliability. Drawing on survey data (n = 148), in-depth interviews, and spatial analysis, the study explores the lived experience of connectivity in Arctic rural communities and introduces a novel \textit{Cellular Coverage Inequality (CCI) Index}. The index combines measures of rurality and network performance to quantify spatial disparities that are masked by national coverage statistics. Results reveal that headline indicators overstate inclusiveness, while local users report chronic connectivity gaps affecting work, safety, and access to services.

Building on these findings, the paper outlines policy reflections in six areas: shared infrastructure and roaming frameworks, spectrum flexibility for rural operators, performance-based Quality-of-Service monitoring, standardized and transparent reporting, temporal and seasonal capacity management, and digital-skills initiatives. Together, these recommendations highlight the need for multidimensional metrics and governance mechanisms that link technical performance, spatial equity, and user experience. The analysis contributes to ongoing debates on how broadband policy in sparsely populated regions can move beyond nominal coverage targets toward genuine inclusion and reliability.
\end{abstract}

\bigskip
\noindent\textbf{Preprint note:} This manuscript has been submitted to \textit{Telecommunications Policy}.



\maketitle

\section{Introduction}\label{sec:1}

Good digital connectivity is fundamental for social inclusion, economic participation, access to digital services, and public safety \citep{WorldBank2016,UnitedNations2022}. However, rural–urban connectivity gaps, large differences in the quality of connectivity between rural and urban regions worldwide, remain a persistent global challenge. Sparsely populated and geographically challenging regions are largely underserved or lack connectivity altogether \citep{Dlamini2021}. In the Arctic region, including northern Finland and Sweden, harsh climate, long distances, and dispersed population patterns complicate the provision of equitable and high-quality broadband \citep{AEC2017}. 

The digital divide has traditionally been conceptualized in terms of availability, namely, whether households or regions are covered by broadband networks. International monitoring frameworks, such as those of the Organization for Economic Co-operation and Development (OECD), often report urban–rural data rate gaps in aggregate terms. Recent OECD figures, for example, show average download data rates in rural areas that lag urban areas by around 20 Mbit/s, underscoring persistent disparities but without revealing how they are distributed spatially or socially \citep{OECD2025}. Conceptual work has since argued for a broader framing. \citet{szabo2024} offers a comprehensive conceptual framework that distinguishes between macro and micro-level inequalities, leading to structural, cognitive, and motivational forms of the digital divide. This perspective highlights that digital inequality emerges from the interplay between structural factors such as infrastructure, policy, and institutional arrangements, and micro-level factors including skills, awareness, and motivation. Empirical research also shows that these patterns are sensitive to governance and policy context. Using U.S. county-level data, \citet{whitacre2020} demonstrate that state broadband policies significantly influence measured availability at the 25/3 Mbit/s benchmark, with funding programmes and active state broadband offices associated with higher rural availability, whereas restrictions on municipal or cooperative provision tend to reduce it. Collectively, these contributions have shifted the understanding of the digital divide toward a multidimensional phenomenon that encompasses social, behavioural, and institutional dimensions alongside physical infrastructure.

At the policy level, the European Union’s (EU) Digital Decade policy programme sets ambitious connectivity goals for the coming years. By 2025, all households should have access to high-speed internet connections of at least 100 Mbit/s, and by 2030, all households should be connected to gigabit-speed networks \citep{EuropeanCommission}. National strategies in both Sweden and Finland align with these ambitions but with different emphases and timelines.

In Sweden, the 2016 broadband strategy, A Completely Connected Sweden by 2025 emphasizes that proper broadband access is essential for participating in society and accessing public services \citep{Sweden_BB_Strategy,Bredbandsforum}. The strategy defines clear progressive targets for 2020, 2023 and 2025, but more importantly, the remaining target for this year states that by 2025 98\% of the households and enterprises should have access to 1 Gbit/s broadband, 1.9\% to 100 Mbit/s, and 0.1\% to 30 Mbit/s. Also, a target for mobile connectivity states that \emph{any place where people normally reside of move} good quality mobile services must be provided. Post- och telestyrelsen (PTS), the Swedish Post and Telecom Authority monitor developments and publishes a yearly report on the progress towards these goals.  In an interpretation of the government's latter goal on mobile services, PTS announced that roughly 4\% of Sweden's total land area constitute \emph{places where people normally reside or move}. In other words, this interpretation excludes 96\% of the land area and specifies and relaxes the goal to only target a fraction of the country. In a general account, PTS's latest monitoring reports suggest that several of these the above goals are unlikely to be fully achieved by 2025 \citep{PTS2024}.

In Finland, the 2019 Digital Infrastructure Strategy set a similar objective: all households should have at least 100 Mbit/s connectivity by 2025, upgradable to 1 Gbit/s in line with the EU broadband strategy \citep{Wiren2019}. Yet official connectivity maps reveal that fixed broadband networks remain unavailable across vast areas \citep{Traficom2025a}. In these areas, Fourth Generation (4G) at 30 Mbit/s covers only 80-95\% of households, and 100 Mbit/s mobile service coverage is considerably lower. For fixed networks, 100 Mbit/s download data rates are available to 81\% of households nationwide, but only to 60\% in rural areas and 45\% in sparsely populated areas \citep{Traficom2025b}. Finnish law mandates a general service requirement under which households and fixed workplaces must have a 4.5 Mbit/s download rate (minimum 3.5 Mbit/s) \citep{Traficom2023,StatuteBookFinland2021}.  While better than no service, this minimum is insufficient for many daily activities, particularly when multiple users share the connection. Furthermore, frequency licenses include the requirement that main roads must be covered. Although Finnish authorities report that nearly 100\% of households have access to 30 Mbit/s 4G \citep{Traficom2024}, this figure masks significant gaps: in some areas, data rates fall below the nominal threshold, and weak coverage is still reported even along main roads and in villages. Despite Finland’s strong overall digital infrastructure, connectivity challenges persist in many Arctic and rural settings. Gaps in mobile coverage, unreliable data connections, and a lack of fiber infrastructure are especially common outside urban centers \citep{Karppinen2024}. Comparable disparities are evident in Sweden’s northern regions, where service coverage and performance vary strongly by geography and settlement density \citep{PTS2024,pts2022}. These national cases exemplify a wider policy challenge common across advanced economies.

These observations demonstrate that even highly digitalized countries face persistent rural connectivity problems, reflecting trends in other parts of the world. Although national agencies have explicitly been tasked with monitoring developments, market investments remain insufficient or incapable of providing connectivity in all places that should be covered. Moreover, the fact that target values are set as percentages of the national population does not help, as this approach completely misses the urban–rural inequality aspects. These weaknesses in the current monitoring and investment frameworks are particularly visible in the Nordic Arctic, where the combination of long distances, low population density, and harsh climate conditions challenges both private and public actors in providing equitable high-quality digital connectivity. 

Building on these challenges, the monitoring of broadband development in Finland and Sweden still relies primarily on headline indicators that inadequately reflect spatial and social disparities. Most official assessments focus on the percentage of the population or households covered by networks that meet a given data-rate threshold. Such measures provide useful national averages but say little about the geographic distribution of coverage or the quality of service experienced by users. In sparsely populated areas, small absolute numbers of connections translate into negligible statistical effects even where coverage gaps are severe. The current practice of expressing targets as population shares therefore tends to overstate inclusiveness and obscure regional inequalities.

Monitoring practices also converge on similar limitations. Indicators that emphasize coverage or data rate can give an incomplete picture of digital access. For example, the International Telecommunication Union (ITU) reports that although global broadband coverage reaches about 96\%, one-third of the world’s population remains offline due to barriers such as affordability and other usage gaps \citep{ITU2024}. At the national level, Sweden’s indicator “fiber in absolute proximity” has been criticized for overstating meaningful access by ignoring installation costs or subscription barriers \citep{slatmo2022}. EU-level monitoring still relies primarily on operator-reported or model-based maps. Under the European Electronic Communications Code, the Body of European Regulators for Electronic Communications (BEREC) requires national regulators to verify these datasets, noting that modelled coverage maps may not accurately represent actual service availability, particularly outside dense urban areas \citep{BEREC2021,ARCEP2020_press}.   Empirical studies further demonstrate the limitations of coarse national indicators. In Finland, \citet{lehtonen2020} show that broadband availability measured at a fine spatial resolution of 1 km $\times$ 1 km reveals much sharper rural inequalities than suggested by national statistics and that improved access correlates with lower depopulation in remote areas. In Ireland, \citet{dempsey2024} find that broadband coverage remains highly polarized, with urban areas enjoying near-complete access while many rural localities are still underserved. They note that common area-based measures can understate rural exclusion, although even alternative methods confirm the persistence of the urban–rural divide. Performance-based evidence from other contexts echoes these findings. \citet{gallardo2024} use speed-test data to show that many of the same socio-demographic factors linked to lower adoption, such as rural residence, poverty, and older age, are also associated with lower data rates. \citet{bauer2022tribal} provide further evidence from the U.S., showing that households in tribal areas not only have lower access rates but also face slower data rates and higher prices than neighboring non-tribal communities. Consequently, official statistics may overstate progress toward universal access and underestimate the persistence of localized weak coverage, especially in rural and sparsely populated regions. The continuing gaps in reliability and affordability indicate that headline indicators capture availability but not the lived quality of service. These discrepancies illustrate how reliance on nominal coverage measures can conceal practical exclusion, particularly in peripheral settlements and along transport corridors.

Persistent measurement gaps highlight the need for analytical tools that more accurately represent the geography and quality of digital connectivity. Population-based statistics capture nominal availability but overlook the complex, spatially uneven realities of network performance in everyday life. Existing frameworks rarely integrate dimensions such as data-rate variability, latency, reliability, or affordability, factors that ultimately define whether connectivity is genuinely inclusive. Approaches that go beyond headline indicators would allow policymakers to identify weak links, compare regions on consistent empirical grounds, and design interventions addressing not only access but also performance and equity. In the Nordic Arctic, where long distances, sparse settlement patterns, and climatic extremes complicate both measurement and investment, there is a particular need for metrics that combine spatial precision with policy relevance.

To address these limitations and respond to this need, this paper makes three main contributions. First, it presents empirical evidence from web-based questionnaires and semi-structured interviews with residents of northern Finland and Sweden, capturing lived experiences of digital connectivity across personal, professional, and community contexts. Second, it introduces a novel digital divide performance indicator that integrates a rurality index with reported mobile coverage quality, providing a spatially grounded measure of digital access inequality. Third, it links these technical and experiential insights to targeted policy reflections, offering practical guidance for narrowing the rural–urban connectivity gap in the Nordic Arctic and providing a transferable framework for other sparsely populated regions.

The paper continues as follows. Section \ref{sec:2} outlines the methodology, describing the study setting, data collection procedures, and analytical approach. Section \ref{sec:3} presents the first set of findings, focusing on how connectivity affects daily life and work settings in northern Finland and Sweden. Section \ref{sec:4} introduces the Cellular Coverage Inequality (CCI) index as a new quantitative tool for measuring the digital divide and assessing areal inequalities in cellular coverage. Section \ref{sec:5} explores forward-looking perspectives on connectivity, including expectations and perceived needs for future digital infrastructure. Section \ref{sec:6} provides policy reflections, linking empirical insights and analytical results to implications for broadband strategy and regulation. Finally, Section \ref{sec:7} concludes by summarizing the main findings and outlining directions for future research.

\section{Methodology}\label{sec:2}
The methodology of this study combined quantitative and qualitative approaches in order to capture both the measurable aspects of connectivity and the lived experiences of residents and organizations. The following subsections first describe the geographic and social context in which the study was carried out (Section~\ref{sec:2.1}), then describe the data collection procedures (Section~\ref{sec:2.2}), and finally present the analytical strategies used to integrate survey data, interviews, and spatial indicators (Section~\ref{sec:2.3}).

\subsection{Study Setting}\label{sec:2.1}
This study was conducted as part of the EU Interreg Aurora project \textit{Arctic 6G} (\url{https://arctic6g.eu}), focusing on the northernmost regions of Finland, particularly Finnish Lapland and adjacent cross-border areas. The research area is characterized by vast geographic distances, low population density, and a strong dependence on mobile and wireless connectivity for essential services, livelihoods, education, and communication. 

\begin{figure}
	\centering
	\includegraphics[width=1.05\textwidth]{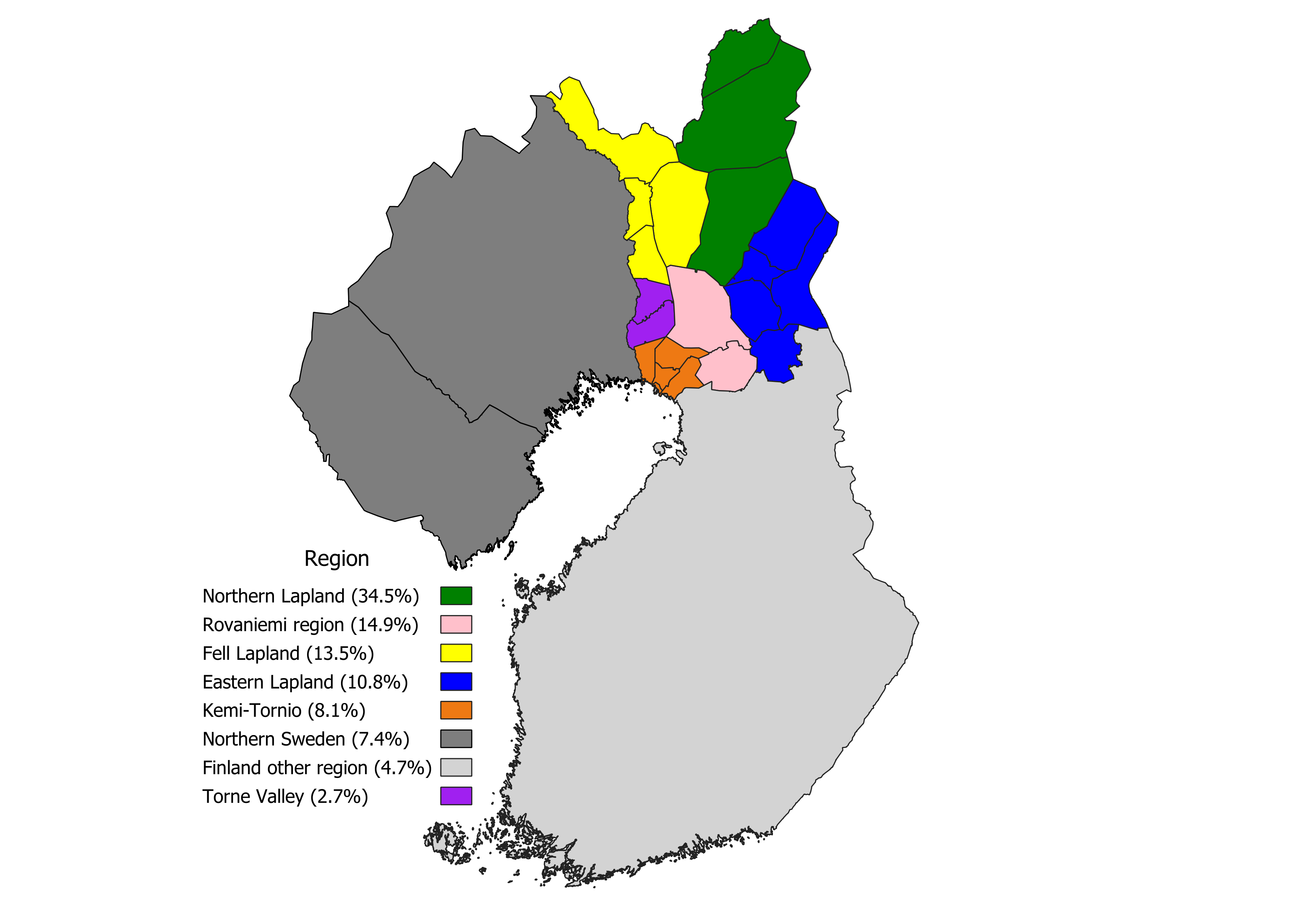}
	\caption{Study area and geographic distribution of survey respondents.}
	\label{FIG:1}
\end{figure}

The geographic scope includes municipalities such as Inari, Sodankylä, Kittilä, and Salla, as well as surrounding regions where tourism, reindeer herding, remote work, and public services all rely on stable internet access. Cross-border contexts with Sweden and Norway are also relevant due to shared economic activities and mobility patterns in the region. The overall study area is illustrated in Figure~\ref{FIG:1}. 

Given the Arctic environment’s extreme climate, seasonal daylight variation, and dispersed settlement patterns, maintaining consistent digital connectivity is both technically and economically challenging. These contextual factors informed the survey and interview design, which sought to capture not only connectivity availability and quality but also the lived experiences of residents, business owners, public sector workers, and students across the region. The following section describes the data collection methods used to achieve this. 

\subsection{Data Collection}\label{sec:2.2}
Data was collected using a mixed-methods approach, combining a structured online survey with in-person and remote interviews conducted during a field visit. This subsection first describes the survey design and respondent demographics, followed by the interview procedures undertaken to complement the survey results. 

\subsubsection{Online Survey}

The online survey was the primary tool for gathering data on connectivity-related experiences, needs, and expectations across Lapland. It was designed to capture usage patterns, challenges, and future requirements in both personal and professional contexts. The instrument included a mix of open-ended and structured questions to generate both quantitative and qualitative data. Three formats of closed-ended questions were used: single-select, multiple-select, and scaled responses. Selected items were followed by open-ended prompts to obtain deeper insight into respondents’ circumstances, capturing the contextual richness and lived experiences behind the quantitative data. 

The questionnaire was available from December 2023 to June 2024 on the Webropol platform in both Finnish and English. It was distributed primarily via email to 1,364 recipients using the Taloustutka Finnish business information service, with additional outreach through the Reindeer Herders’ Association’s contact lists. To maximize visibility, the survey was also published on the 
\href{https://www.ltu.se}{Luleå University of Technology (LTU)}, 
\href{https://www.lapinamk.fi}{Lapland University of Applied Sciences (Lapland UAS)}, 
\href{https://www.oulu.fi}{University of Oulu (UOulu)}, and the 
\href{https://arctic6g.eu}{Arctic6G project} websites, 
and promoted through social media platforms including 
LinkedIn and Facebook.

A total of 148 individuals responded to the survey, representing a range of age groups, industries, geographic locations, and professional sectors. The geographic distribution of respondents is shown earlier in Figure~\ref{FIG:1}. The largest share (34.5\%) came from Northern Lapland (Sodankylä, Inari, Utsjoki), followed by the Rovaniemi region (14.9\%), Fell Lapland (13.5\%), Eastern Lapland (10.8\%), Kemi–Tornio (8.1\%), Northern Sweden (7.4\%), other Finnish regions (4.7\%), and the Torne Valley (2.7\%). No responses were recorded from Northern Norway. 

\begin{figure}
	\centering
	\includegraphics[width=.9\textwidth]{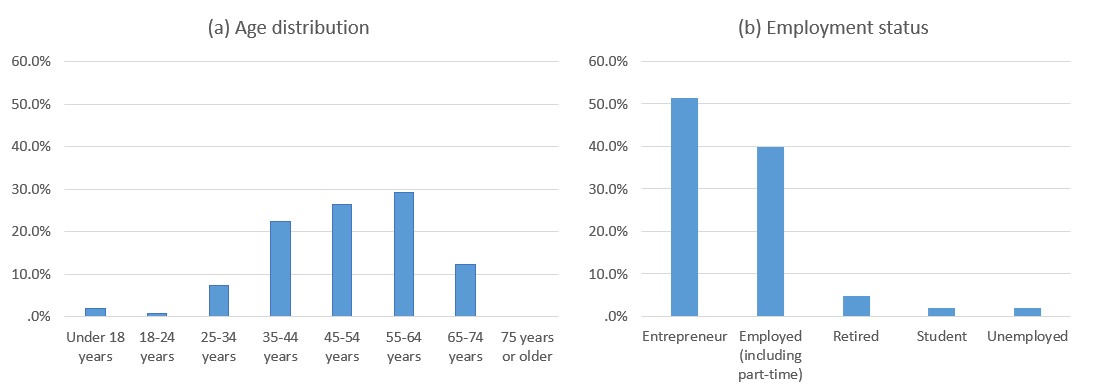}
	\caption{Respondent demographics: (a) age distribution, (b) employment status}
	\label{FIG:2}
\end{figure}

The age profile was weighted toward working-age adults, with the largest groups being 55–64 years (29.1\%), 45–54 years (26.3\%), and 35–44 years (22.3\%). Younger adults aged 25–34 accounted for 7.4\% of respondents, while 18–24-year-olds represented 0.7\% and those under 18 made up 2.0\%. No participants were aged 75 or older. Just over half of respondents (51.4\%) identified as entrepreneurs, while 39.9\% were employed either full-time or part-time. Smaller shares included retirees (4.7\%), students (2.0\%), and unemployed individuals (2.0\%), as shown in Figure~\ref{FIG:2}.
In terms of occupational background, respondents represented diverse sectors. The most common fields were tourism (20.3\%), education (18.6\%), reindeer husbandry (8.5\%), and trade (5.1\%), with the remainder spread across technology, healthcare, social and rescue services, infrastructure, mining, public administration, and information and communication technology (ICT). Among employees, organization sizes were relatively evenly distributed, with notable shares in both small organizations (10–49 employees, 27.6\%) and large organizations (250+ employees, 25.9\%). A smaller proportion (20.7\%) worked in micro-enterprises with fewer than 10 employees. Among entrepreneurs, 89.3\% operated micro-enterprises with fewer than 10 employees, and only 8.0\% ran businesses with 10–49 employees. These patterns indicate that both employees and entrepreneurs are strongly concentrated in micro- and small enterprises, with only a smaller share working in large organizations. Supplementary research illustrations can be found in \citet{Ahvenjarvi2025}'s master’s thesis, which provides additional visual context to some of the results presented in this article.

\subsubsection{Field Visits and Interviews}
As part of the Arctic 6G project’s data collection activities, a follow-up field visit was organized in December 2023 to Northern Finland, covering the regions of Salla, Ivalo, Inari, and Kittilä. This built on a previous visit to municipal government and public safety actors in Lapland, whose results were reported in \citet{Karppinen2024}. During this fieldwork, researchers conducted semi-structured interviews and group discussions with local stakeholders to better understand connectivity challenges in Arctic and rural settings. In total, approximately 67 individuals participated across multiple contexts, including educators and students, tourism entrepreneurs, representatives of traditional livelihoods, local administrators, and community broadband advocates. Sessions were conducted in a mix of group interviews, individual interviews, and small group discussions. Researchers took detailed field notes to capture participants’ experiences and needs, focusing on network reliability, impacts on safety, education, tourism, remote work, and expectations for future technologies. Table~\ref{tab:field_interviews} summarizes the interview contexts and participant groups.

\begin{table}[htbp]
\centering
\caption{Summary of field interviews and participant groups (December 2023 field visit)}
\label{tab:field_interviews}

\begin{adjustbox}{max width=\textwidth}
\begin{tabular}{p{3.2cm} p{4cm} p{7cm}}
\toprule
\textbf{Context} & \textbf{Interview type} & \textbf{Participants} \\
\midrule

Education sector & Two group interviews &
\makecell[l]{%
approx.\ 60 total across two schools:\\
Group 1 – 20 (5 staff, 15 students)\\
Group 2 – 40 (10 staff, 30 students)
} \\

\addlinespace

Rural administration & One phone interview &
1 local community representative \\

\addlinespace

Local tourism & Three face-to-face interviews &
2 tourism entrepreneurs, 1 business manager \\

\addlinespace

Traditional livelihoods & Two face-to-face interviews &
2 representatives from local herding community \\

\addlinespace

Local broadband & One face-to-face interview &
1 community fiber advocate \\

\bottomrule
\end{tabular}
\end{adjustbox}

\end{table}

\subsection{Analytical Approach}\label{sec:2.3}
Quantitative data from closed-ended items were processed using descriptive statistics to identify general patterns in connectivity usage, service quality, and perceived needs, with comparisons made across occupational and regional groups where relevant. 

Responses to open-ended questions were analyzed using thematic review. Given the straightforward nature of these answers, formal coding was not deemed necessary. Instead, common themes were identified through direct content review and comparison of response patterns. This process allowed qualitative findings to retain the language and emphasis of respondents, preserving the contextual richness of their lived experiences. The thematic analysis was organized into two main domains: (i) daily life context and (ii) occupational context. To ensure consistency and minimize potential bias, the thematic interpretations were cross-checked against detailed field notes from the in-person interviews conducted during the December 2023 field visit. Two additional researchers independently reviewed the thematic summaries, confirming the stability and clarity of the categorization.  

In addition to this qualitative and survey-based analysis, a spatial performance measure was developed to systematically capture the geography of the digital divide. This involved calculating a \textit{Rurality Index} that assigns values of remoteness to each grid cell in the study region based on distances to settlements of five different population thresholds (200, 1,000, 3,000, 30,000, and 60,000 inhabitants). The rurality scores were then combined with official 4G and 5G coverage data from national regulators to produce a \textit{CCI index}. This key performance indicator (KPI) highlights areas where rurality is high, but network performance is low, providing a spatially grounded complement to the citizen-reported experiences analyzed earlier.  

The following sections present the results of this analysis. Section~\ref{sec:3} reports how connectivity availability, quality, and reliability shape daily life and work in the study area, while Sections ~\ref{sec:4} and ~\ref{sec:5} extend the discussion toward spatially grounded performance measures and future expectations, respectively.  

\section{Connectivity Impacts in Daily Life and Work Settings}\label{sec:3}

This section presents findings from the online survey and supplementary field interviews, focusing on how connectivity availability, quality, and reliability shape daily life and work in the study area. The results are organized into two parts: Section~\ref{sec:3.1} examines impacts on personal and household contexts and Section~\ref{sec:3.2} addresses occupational settings. Both quantitative survey data and qualitative insights from open-ended responses and interviews are integrated to provide a comprehensive view of the lived connectivity experience in Arctic and rural northern regions.

\subsection{Connectivity in Daily Life}\label{sec:3.1}
Survey and interview results show that connectivity has become a central part of daily life in northern Finland and Sweden. People depend on it not only for communication and entertainment but also for tasks such as banking, education, and accessing public services. Before turning to the challenges that arise from poor coverage or unreliable connections, it is useful to understand how people actually use the internet in their day-to-day lives. The following subsection outlines usage patterns, including frequency, timing, and typical applications.
\medskip

\subsubsection{Patterns of Internet Use and Connectivity Disparities}\label{sec:3.1.1}

\begin{figure}
	\centering
	\includegraphics[width=.9\textwidth]{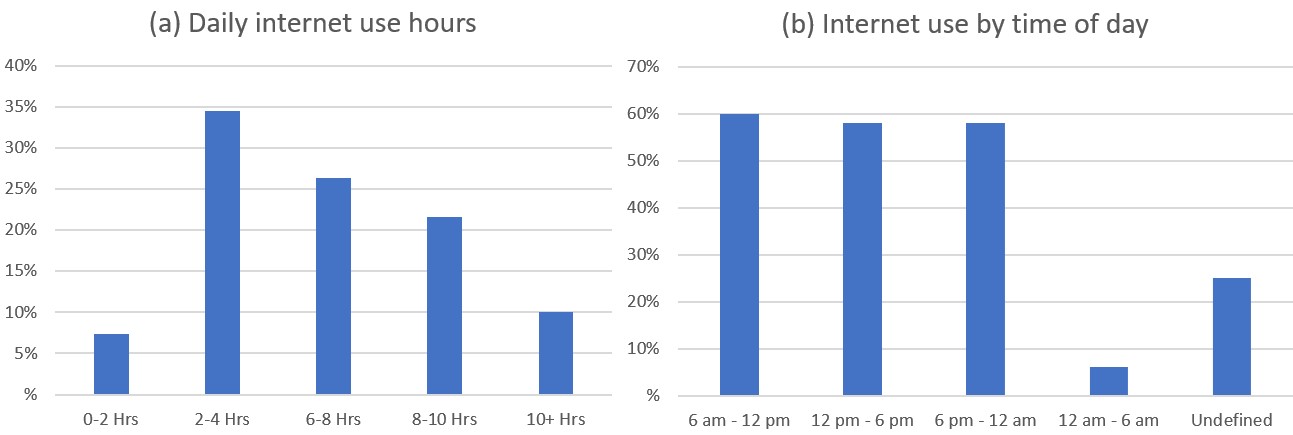}
	\caption{Internet usage patterns: (a) daily hours, (b) time of day. Note: Percentages exceed 100\% for time-of-day usage because respondents could select multiple options.}
	\label{FIG:3}
\end{figure}

Internet connectivity is deeply embedded in the routines of respondents. More than half of respondents used internet-dependent services for over six hours per day, with 26\% spending 6–8 hours online and 22\% spending 8–10 hours. A further 10\% reported more than 10 hours of daily use, while only 7\% used the internet for fewer than two hours. Usage was spread across the day, with 60.1\% active in the morning, 58.1\% in the afternoon, and 58.1\% in the evening, while only 6.1\% used the internet between midnight and 6:00 a.m. Because respondents could select multiple time periods, percentages for time-of-day usage exceed 100 (Figure~\ref{FIG:3}).

\begin{figure}
	\centering
	\includegraphics[width=.9\textwidth]{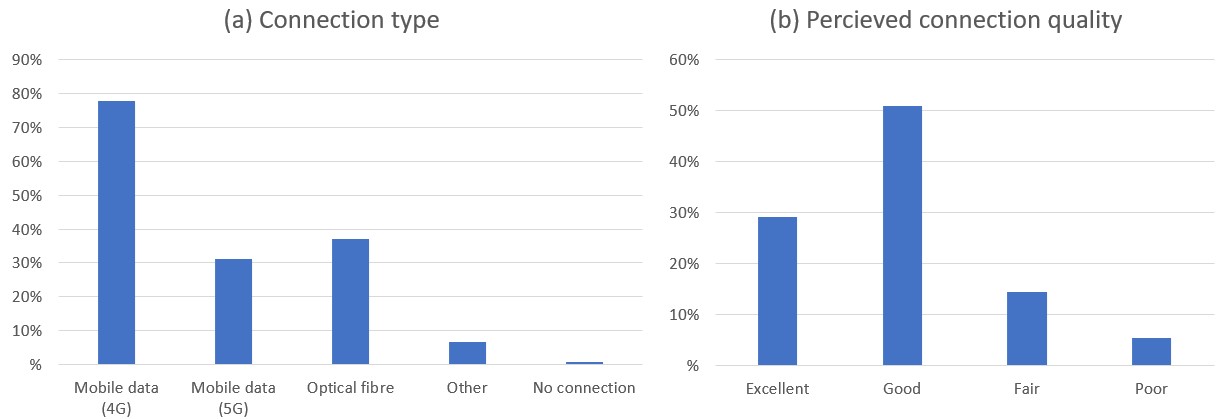}
	\caption{Types of internet connections (multiple responses allowed) and self-rated residential connectivity quality. Note: Percentages for connection types exceed 100\% because respondents could select multiple options.}
	\label{FIG:4}
\end{figure}

Respondents relied on a mix of connection types and could select more than one. A large majority (77.7\%) used 4G mobile connections, 31.1\% reported 5G, and 37.2\% had optical fibre at home. Smaller shares used other technologies such as Starlink, 3G, or fixed 5G, and only one respondent reported they had no internet connection. Most rated their residential connectivity positively, with 50.9\% describing it as “good” and 29.1\% as “excellent.” However, 14.5\% rated it “fair” and 5.5\% as “poor,” indicating that quality remains uneven (Figure~\ref{FIG:4}).

Survey responses highlighted multiple activities that require fast and reliable internet (multiple selections allowed). Work‑related functions were most common: video conferencing/virtual meetings (74.3\%), cloud‑based services (64.2\%), and downloading/sending large files (55.4\%). Frequently cited personal and mixed uses included high‑definition video/movie streaming (46.6\%) and real‑time navigation/mapping (43.2\%). IoT devices (e.g., smart home automation/connected devices) were used by 27.0\%, and online gaming by 12.2\%. An additional 13.5\% provided “Other” uses, which included emergency calls, payment processing and invoicing, cash registers/booking systems, remote monitoring of agricultural equipment (e.g., milking robots), programming, customer communications, public‑service/banking transactions, mobile phone/TV use, and remote Sámi‑language instruction with continuous video linking to pupils. This question allowed multiple selections, so percentages sum to more than 100\%.

A significant portion of respondents (91.9\%) reported spending time in sparsely populated or wilderness areas, such as cottages or natural settings. More than half (61.8\%) reported that they experience different internet needs or challenges in these locations compared to their home or workplace. In fact, an overwhelming majority (95.3\%) had noticed a clear difference in service quality between populated areas and remote regions. When asked about the importance of consistent internet performance in both urban and rural settings, 67.6\% rated it as very important, 27\% as important, and 5.4\% as slightly important. No respondents considered it not important at all.

These findings highlight that spatial disparities in connectivity are not limited to permanent residences but extend into transitional and rural spaces where people live, work, and travel. In northern contexts, reliable internet access is increasingly essential for participation in digital society and for maintaining everyday safety and wellbeing. Thematic analysis of open-ended responses, presented in the next subsection, deepens this understanding by examining how such disparities are experienced in everyday life.

\subsubsection{Lived Experiences and Challenges}\label{sec:3.1.2}

While the quantitative results highlight widespread use and reliance on multiple connection types, the open-ended responses reveal important nuances in how these connections function in practice. Thematic analysis on the open-ended responses uncovered four key areas of concern: frustrated online experiences with interruptions and slow data rates, near-complete lack of coverage in rural areas, poor connectivity affecting leisure activities, health and safety risks.

\newcommand{\theme}[1]{\vspace{1em}\noindent\textit{#1}\par}
\theme{Frustrated Online Experiences due to Interruptions and Slow Data Rate}

Participants frequently described situations where even basic online tasks were disrupted. Online banking, form submissions, and email use could be derailed by connection drops, often at the final step, forcing the entire process to be repeated. As one rural resident explained: \textit{“Nowadays, many things have to be done online, and it is really frustrating when you have filled out forms/documents and during the submission phase you get a comment saying, ‘an error occurred,’ your data was not saved, and you have to start over.”}

Underlying these experiences was a strong sense of resentment toward what participants perceived as a widening gap between marketing promises and real-world delivery. As one participant noted: \textit{“Current mobile phones are about 10 Mbit/s instead of the promised 300 Mbit/s.”} Interviewees in Ivalo and Inari reinforced this view, describing how they had been sold 5G subscriptions despite living in areas with no functioning 5G. In practice, they found that phones often worked better when the 5G setting was disabled. One respondent summarized the mismatch with a sharp metaphor: \textit{“For the customer, it’s like being sold a Ferrari but getting a moped.”} Such frustrations illustrate not only dissatisfaction with actual performance but also growing mistrust toward operators and doubts about the value of future network generations.

\theme{Near-complete lack of coverage in rural areas}

The situation was consistently described as worse in remote villages and sparsely populated zones. In certain parts of Sodankylä, mobile networks were said to “hardly work at all,” with even SMS and calls often failing. One resident put it bluntly: \textit{“In the more remote villages of Sodankylä, the mobile network hardly works at all. Normally sending/receiving text messages and making calls is almost impossible.”}

In towns like Sodankylä and Kittilä, 4G networks could collapse during evening peak hours, leaving residents cut off during critical times. In such places, fiber, when available, became the only option. Field interviews confirmed similar problems elsewhere: participants at a regional education institute reported that already five kilometers outside the village, 4G has become unreliable, making it difficult for students to attend remote classes or complete schoolwork from home. Cottage owners in Salla were unable to work remotely due to unstable connections, despite increasing expectations for remote work.

Border regions presented distinctive challenges. In interviews near Lake Inari, respondents explained that their phones frequently switched to Russian networks, forcing them to turn devices off altogether to avoid roaming charges or unreliable service. Reindeer herders described similar problems along the Norwegian border, where calls would unexpectedly connect through Norwegian operators. Although roaming within Norway is cost-free under the EU/European Economic Area (EEA) rules \citep{EU_Roam_2022,EEA_JC_2022}, the unintended network switching still disrupted communication and reduced reliability. These experiences illustrate how fragile coverage in frontier areas is compounded by cross-border interference, with differing implications depending on the neighboring country.

\theme{Poor Connectivity Affecting Leisure Activities}

Connectivity issues followed people into their leisure time, disrupting activities that ranged from streaming television to listening to radio, audiobooks, or podcasts. Some respondents said TV channels only worked intermittently, while others noted that outdoor activities were compromised by lost of updating online maps in the smart phone (and sometimes Global Positioning System (GPS) black spots); that means losing navigation in remote terrain made hiking, sledding, or off-roading riskier. This suggests that in addition to network improvements, digital skills such as knowing how to use offline map applications could help mitigate risks in remote areas. A recurring frustration was the inability to connect from cottages, particularly for those who wanted to work remotely or support children’s studies during holidays. As one participant noted: \textit{“The connections at the cottage don’t work at all. I could do, for example, accounting work there, and the children could do things related to their studies.”} 

Field interviews reinforced these patterns. Young people in Ivalo pointed out that connection prices were high compared to the quality, making them feel excluded from everyday digital entertainment. Others described how even simple leisure activities, such as streaming a film on Apple TV, became impossible in the evenings when demand peaked. It should be noted, however, that these problems were not universal: popular skiing resorts and other well-covered tourist destinations generally offered reliable service, while difficulties were more commonly reported from remote cottages and sparsely populated areas.

\theme{Health and Safety Risks}

Connectivity problems were repeatedly linked to serious safety hazards. In some cases, a simple navigation failure in the wilderness could escalate when weather, terrain, or vehicle breakdowns left individuals without the means to call for help. Residents described being stranded in such situations, whether on icy roads, deep in the wilderness, or along uninhabited shores. Even when emergency assistance was available, paramedic operations faced critical challenges due to poor network conditions, including disrupted access to patient information systems and delays in sending data for medical consultation. These disruptions were seen as posing dangerous risks to health and safety. A paramedic working in a remote area explained: \textit{“We use an electronic patient information system in real time. Many times, when consulting a doctor, you cannot get all the data on the doctor's screen because the communication connections are not good enough. This can, at worst, cause dangerous situations that are harmful to health.”}

Field interviews highlighted further risks related to everyday navigation. Interviewees in Ivalo described how glitching map applications on snowmobiles created life-threatening risks in snowstorms or darkness: \textit{“Map applications should function reliably, but at present they glitch while moving and create risks, for example, when going off route and potentially driving a snowmobile off a cliff.”} Similar challenges occur on Lake Inari, where boating navigation systems can drift up to 300 meters off course, raising concerns about accidents. These accounts show that unreliable connectivity not only disrupts daily routines but also undermines trust in safety-critical digital applications.

\medskip
These four themes, while varied in their specific impacts, share a common underlying factor: connectivity quality is highly dependent on location, with service reliability often declining sharply outside populated areas. The interviews further show that residents not only face technical disruptions but also encounter mistrust, digital exclusion, and heightened risks to safety in their daily lives.

\subsection{Connectivity in Work Settings}\label{sec:3.2}
The data indicate that connectivity plays a central role in shaping everyday work practices across the study area. Stable and high-quality internet access has become an essential infrastructure for communication, transactions, and the use of digital tools in both private and public organizations. This dependency extends beyond conventional office environments to include field-based and traditional occupations. The following analysis draws primarily on quantitative survey data to examine the types of connected devices used by individuals and organizations, their operational purposes, and the varying levels of digitalization across business functions. These results are complemented by a qualitative examination of lived experiences and challenges within work context, presented in Section~\ref{sec:3.2.2}.

\subsubsection{Digital Adoption and Connectivity Dependence}\label{sec:3.2.1}

\begin{figure}
	\centering
	\includegraphics[width=.9\textwidth]{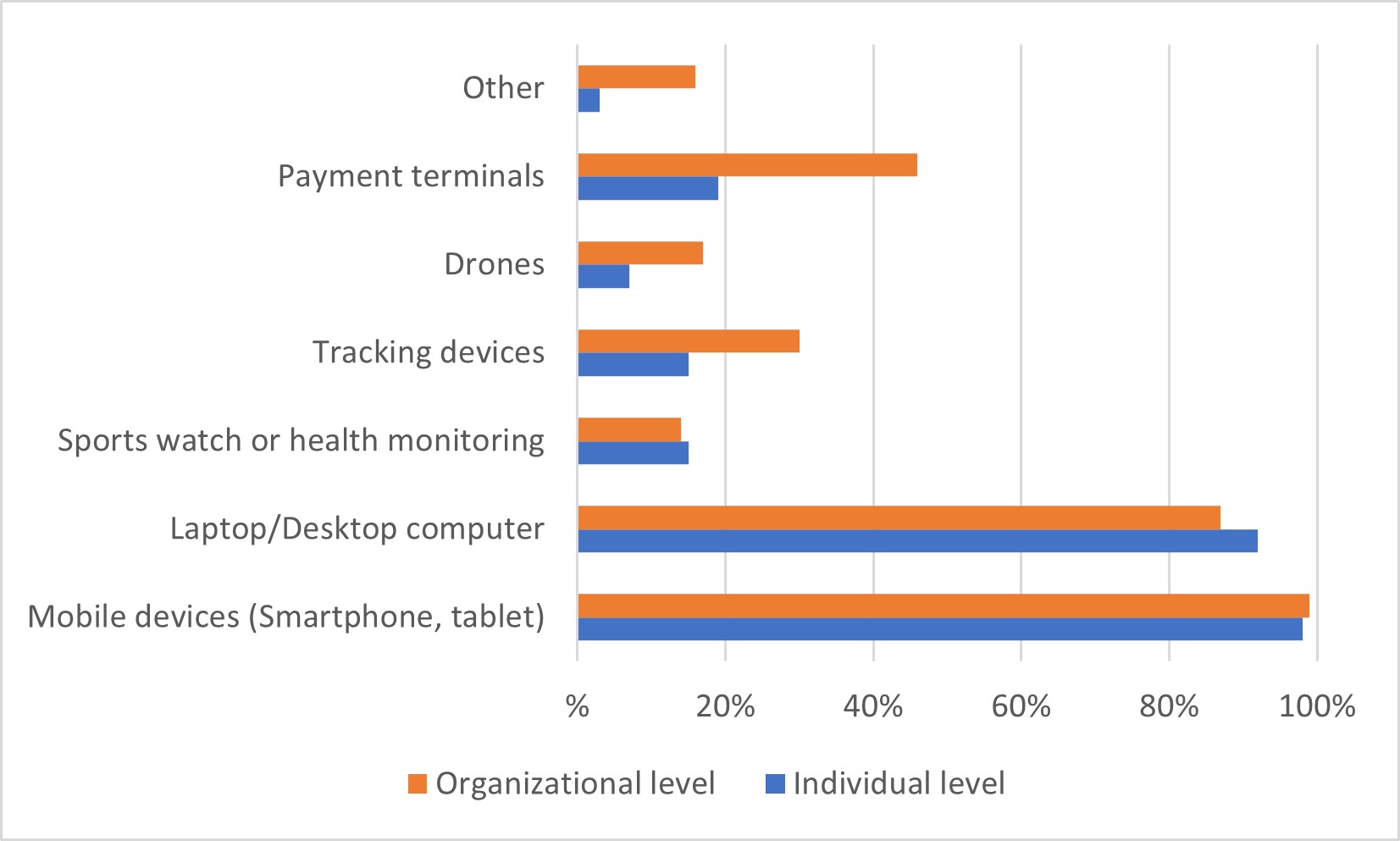}
	\caption{Usage of connected devices at individual and organizational levels. Note: multiple selections allowed.}
	\label{FIG:5}
\end{figure}

Workplace operations across the study area show a high degree of dependence on internet-connected devices. Figure~\ref{FIG:5} illustrates the survey responses (multiple selections allowed) regarding the types of internet-connected devices individuals use in their work and those utilized by their organizations. At the individual level, 98.0\% of respondents reported using mobile devices (smartphones or tablets) for work, while 92.0\% used computers (laptops or desktops). At the organizational level, mobile device use was nearly universal (99.0\%), with computers used by 87.0\%. In addition to this, 46.0\% of companies reported using payment terminals, 30.0\% used tracking devices, and 17.0\% incorporated drones into their operations. Sports watches or other health-monitoring devices were reported by 15.0 \% of individuals and 14.0\% of companies.

Other reported workplace devices included alarm extensions, forestry machinery, industrial equipment, surveillance systems, and digital tools for managing employee and customer information. Only one respondent indicated that their work did not require internet connectivity, citing a complete lack of network access in remote reindeer herding locations. All others confirmed that connectivity was essential to their business operations.

Beyond device usage, survey responses also highlighted how organizations depend on internet connectivity for specific operational needs, including large file transfers, remote system access, and cloud-based services. 72\% of the respondents reported that their companies require connectivity to transfer large image files and video files in their operations. However, only 43\% of the companies required remote access to information systems such as CRM (Customer Relationship Management), ERP (Enterprise Resource Planning) systems, or internal databases. 61\% of the respondents reported that their companies are already utilizing cloud-based implementations in their internal processes, while 13\% have plans to use these services in their internal processes in the future.

When asked to identify the industries that would most benefit from improved internet connectivity in northern Finland, respondents most frequently selected healthcare (68.9\%), followed by education (54.7\%) and tourism (54.7\%). Healthcare was seen as the highest priority, reflecting the critical role of reliable networks for telemedicine, remote diagnostics, and digital patient management in sparsely populated regions. Emphasis on education reflects its reliance on stable internet for remote learning and digital teaching tools, particularly in areas where distance education is essential. As previously identified, the tourism sector is dependent on functioning networks for bookings, payment services, and ensuring safety and communication in remote destinations.

While healthcare, education, and tourism emerged as the top three sectors, respondents also identified several other fields where improved connectivity could have a meaningful impact. Leisure activities (28.4\%) illustrate the growing integration of digital services into recreation and outdoor experiences, while reindeer husbandry (20.9\%) increasingly depends on remote tracking technologies. Industry, including forestry, mining, sawmilling, and construction (17.6\%), along with transport and road maintenance (16.9\%), were also identified as areas where digital infrastructure could enhance operational efficiency and safety. Smaller but notable shares of respondents selected entertainment and media (15.5\%) and smart city infrastructure (12.8\%). An additional 9.5\% of respondents specified “other” sectors, most commonly referring to agriculture, cultural activities, Information Technology (IT) and remote work, and public services.

Overall, connectivity is viewed as essential across diverse sectors but is most strongly associated with healthcare, education, and tourism. Digital infrastructure is thus seen not only as a general-purpose enabler but as a prerequisite for maintaining welfare services and enhancing the competitiveness of key regional industries.

\begin{figure}
	\centering
	\includegraphics[width=.9\textwidth]{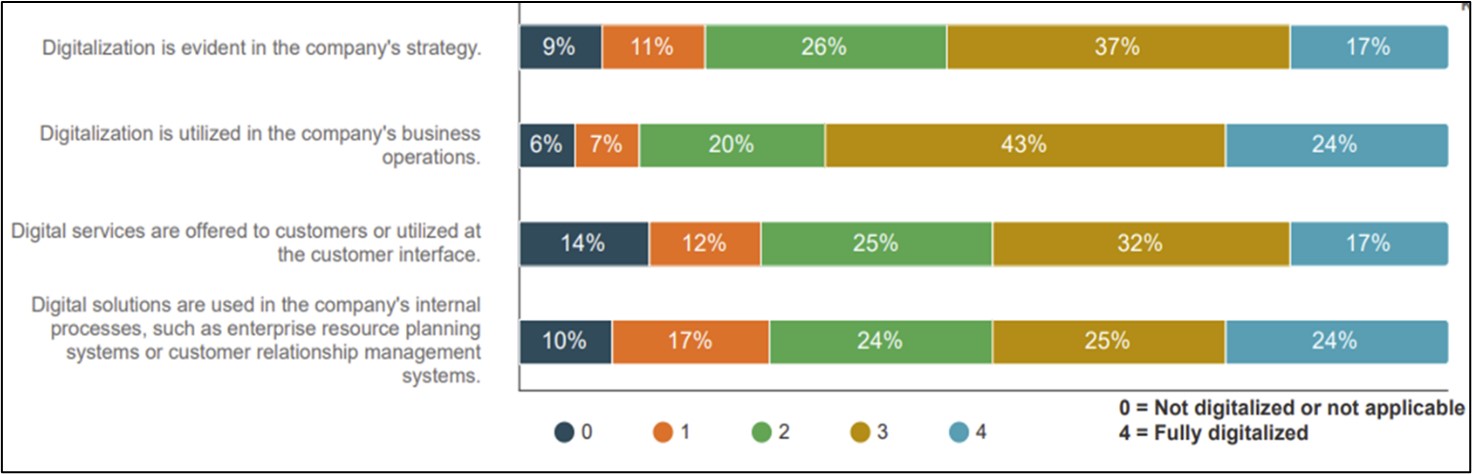}
	\caption{Digitalization level of business functions}
	\label{FIG:6}
\end{figure}

While connectivity enables digital activity across sectors, the degree to which organizations have adopted digital practices internally varies. To better understand this, respondents were asked to evaluate the level of digitalization across four business functions: company strategy, business operations, internal processes, and customer interaction, with a 5-point scale where 0 being the lowest and 4 being the highest in terms of digitalization. Results are summarized in Figure~\ref{FIG:6}.

Results indicate a varied but generally positive trend of digitalization within business operations. 54\% recognized the strategic aspect at levels of 3 or 4, indicating a high recognition of digitalization as an essential part of corporate strategy. Business operations show an even stronger digital presence with a 67\% rating at levels 3 and 4. This is an indication of a higher level of integration of digital tools into core business activities. Internal digital solutions, such as ERP and CRM systems, are moderately adopted: 49\% rated their use at levels 3 or 4, while 27\% rated them at 1 or 2, suggesting that internal digitalization remains in progress. A very similar trend was reported in customer interactions, with 49\% rating at higher levels (3 and 4) while 26\% rating at lower levels (0 and 1). This indicates that even though businesses are getting better at providing digital services, digital customer interaction strategies still need to be improved. These findings highlight that while digitalization is being implemented across various dimensions, there remain opportunities for companies to enhance their digital capabilities, more specifically in customer engagement and internal processes.

Overall, the survey findings indicate a broad but uneven level of digital adoption across sectors, underscoring the growing importance of connectivity in business operations. However, this growing dependence also introduces significant challenges when networks are unstable or unavailable. To complement the quantitative results, a thematic analysis of open-ended responses was conducted. The following section presents the main themes identified, illustrating how connectivity-related disruptions affect everyday work practices and organizational performance.

\subsubsection{Lived Experiences and Challenges}\label{sec:3.2.2}
While the quantitative findings highlight the importance of connectivity for business functions, the open-ended survey responses reveal the vulnerabilities that arise when networks are unstable or unavailable. To deepen understanding of these issues, a thematic analysis was conducted to identify recurring patterns in respondents’ experiences. Four main themes were identified: disruptions to remote work, challenges in maintaining customer service and experience, reduced operational efficiency, and concerns related to occupational health and safety.

\theme{Hampering Remote Work}

Connection drops, low data rates, and lag of connectivity were frequently cited as barriers to effective remote work. Respondents reported interruptions to online meetings, with freezing or grainy video feeds undermining communication. These issues extended to remote server access, slowing or preventing work tasks outside the office. One participant described how location dictated work feasibility. As one participant explained: \textit{“The network works well in my own area. However, I was working in Ivalo, where even 3G was in use at times. It made it very difficult to do work.”} Such variability forced workers to avoid certain locations for remote work, even if their job tasks were otherwise location independent.

Field interviews reinforced these challenges. One participant explained that when an employer’s operator failed, work calls had to be made on a personal subscription, and some people even bought multiple subscriptions to stay connected. Although these measures allowed work to continue, they are inefficient and costly, highlighting how weak infrastructure forces people to bear the burden of poor connectivity.

\theme{Hindering Customer Service and Experience}

Connectivity failures directly affected service delivery and client interaction. Payment terminals sometimes dropped mid-transaction, leading to incomplete payments and dissatisfied customers. Slow-loading websites and lost online form data created delays in responding to client requests. Poor or intermittent mobile signals disrupted voice calls, affecting both initial inquiries and ongoing communication. These challenges were especially problematic in the tourism sector, where businesses depend on reliable communication with visitors. As one tourism operator recounted: \textit{“I have a remote tourist destination, and my clients know that the internet doesn't work properly there, but if you need it, you have to hike up a mountain for a while. I've considered putting an antenna on a higher tree or something.”} 

Field interviews revealed similar problems. A participant from tourism sector explained that bookings, cash registers, and guide-tracking applications all relied on stable internet connections. When the network failed, these systems collapsed in a “chain reaction” that disrupted both staff and customers: \textit{“If a receipt can’t be printed, sales cannot be completed.”} Such accounts underline how connectivity problems not only slow service but also erode customer trust and threaten the reputation of businesses operating in remote areas.

\theme{Reducing Operational Efficiency}

Connectivity disruptions stalled routine processes such as invoicing, updating inventory systems, and software configuration. Field-based industries, including forestry and agriculture, often required staff to travel off-site to upload data or access online tools. Seasonal congestion during peak tourism or production periods made real-time data transfer, such as machine monitoring or reindeer tracking, nearly impossible. Warehouses without consistent internet connections faced similar difficulties, as online instructions or simulations became unreliable. One respondent highlighted this challenge: \textit{“The company's warehouse, i.e. the premises, should have an uninterrupted internet connection so that it can use the internet instructions and simulations when we configure the devices. Speed 50} Mbit/s \textit{would be enough, but the connection drops.”} 

Field interviews added to these accounts. Tourism entrepreneurs near Ivalo described how connections sometimes cut out so badly that even basic services such as Google would not load. They also reported that reservation platforms like booking.com frequently failed during submission, forcing staff to re-enter information and causing delays in confirming customer bookings. Others explained that equipment upgrades, such as new routers or antennas, often brought little improvement, creating additional costs without solving the underlying issue of weak infrastructure. These experiences show how unreliable connectivity wastes staff time, increases workloads, and adds hidden costs that undermine overall business efficiency.

\theme{Occupational Health and Safety Concerns}

Connectivity failures created serious, sometimes life-threatening, safety risks in occupational settings. Respondents described situations where injured employees, ill clients, or drivers involved in accidents could not call for help because mobile coverage was absent. In remote industries such as forestry, agriculture, and tourism, the inability to transmit location data or communicate in real time meant that minor incidents could escalate rapidly. Supervisors reported losing contact with teams for hours, forcing them to suspend work and travel long distances to physically locate missing staff. One participant explained: \textit{“For example, in forestry work, you have to interrupt work and drive the car to the coverage area. Or, you have to spend hours of your working time going outside the coverage area to check why the employee has not arrived from the forest. Poor coverage is a problem for ensuring occupational safety.”} Such gaps not only endanger workers but also undermine compliance with occupational safety standards and increase employer liability. The findings point to a systemic vulnerability: in large parts of rural Lapland, safety-critical communication depends on network infrastructure that is not designed to guarantee coverage in all work locations.

Field interviews highlighted similar vulnerabilities in traditional livelihoods. Members of a reindeer herding cooperative recounted an incident in which, during a winter medical emergency at their cabin, only one window provided mobile coverage, which allowed WhatsApp but not regular calls, so the phone had to be left there while communication continued through a Bluetooth headset. The case illustrates how inadequate connectivity can turn already difficult situations into life-threatening ones, underscoring the importance of reliable networks for occupational safety in remote environments.

\medskip
Taken together, these four themes highlight how business operations in Arctic and rural regions are constrained less by ambition than by location. Connectivity gaps undermine remote work, service delivery, efficiency, and occupational safety, with consequences ranging from lost productivity to life-threatening emergencies. Field interviews underline that workers and enterprises are forced into costly workarounds, face reputational harm when services fail, and encounter risks when digital tools cannot be relied upon. As with daily life, the core issue is not the absence of digital needs or aspirations, but the uneven quality and reliability of networks across places.

\section{Measuring the Digital Divide}\label{sec:4}

Building on the findings from Section~\ref{sec:3}, which documented the lived impacts of poor connectivity in daily life and work, this section turns to the challenge of measurement. For years, rural and remote areas have struggled with inadequate or poor-quality cellular coverage. Cellular networks typically launch new technologies, like 5G, first in urban areas due to high user density and cost-effectiveness. Once coverage is established in cities, new networks gradually expand to rural areas -- an expansion that often takes several years.

Operators and regulators typically express cellular coverage as either the percentage of the area or the percentage of the population, that is covered by a cellular network, and on a national basis.  The \emph{Areal} Coverage Ratio (ACR) is defined as the percentage of total \emph{area} within a country that has cellular coverage. While a high ACR is desirable for assessing overall coverage, it does not provide insights into how coverage is distributed across different areas or how it has expanded over the years. Specifically, it would be important to analyze whether coverage has increased more in urban areas or in rural regions, as this information can significantly impact service accessibility and user experience.  Meanwhile, the \emph{population} coverage ratio (PCR), which operators also use, expresses the percentage of the \emph{population} that has access to cellular coverage based on where they live. However, this measurement also has its limitations; it does not indicate areas with no coverage. Furthermore, people are often distributed sparsely in rural areas, while a significant portion of the population resides in urban areas. Additionally, this method relies on registered addresses and does not account for where people work, their summer homes, or the roads they travel. As a result, it may not accurately reflect the true accessibility of cellular coverage for individuals in various contexts.

In short, these figures do not accurately reflect fairness of cellular coverage, as they fail to account for how coverage has expanded over time or for the socio-economic and geographic status of uncovered areas. High levels of urban coverage can mask persistent gaps in sparsely populated regions, where access to connectivity is often most critical for inclusion, safety, and economic participation.  To investigate this inequality, we review here the recently developed CCI index, see \citep{CCI25}. This CCI index illustrates the distribution of coverage among people with varying levels of rurality, providing a clearer picture the \emph{distribution} of cellular coverage, of the disparities in access. 

Computing the CCI index for a certain country or geographical region requires availability of two distinct maps of the region. The first map is a \emph{rurality map} of the region. It typically uses metrics such as population and distance from urban centers. The second map is a cellular \emph{coverage map} of the region that shows where cellular services are provided and where they are not. By combining these two maps, we can then construct a \emph{cellular coverage concentration curve}, from which the CCI index is calculated.

In the following subsections, we discuss how to generate and create a rurality map (Section~\ref{sec:4.1}), a cellular coverage map (Section~\ref{sec:4.2}), and in (Section~\ref{sec:4.3}), we apply the CCI index to four study areas: Finland (national), Sweden (national), the Arctic region of Finland, and the Arctic region of Sweden.

\subsection{Rurality Map}\label{sec:4.1}
Defining rurality and an associated rurality map is difficult. A number of measures have been proposed that vary depending on a country's population, geographical features, and environmental conditions. The \emph{Degree of Urbanisation (DEGURBA)}, for instance, is a European statistical classification system that situates geographic areas on an urban-rural continuum \citep{EU2021}. It categorizes regions into three distinct classes: cities (densely populated), towns and suburbs (intermediate density), and rural areas (thinly populated) based on population size, density, and the geographical contiguity of populations within 1 km² grid cells. However, it is important to note that EU countries have different population structures and geographical characteristics. For instance, in Sweden, Norway, and Finland, fewer people live in Arctic areas, primarily due to harsh weather and environmental factors such as long winters. Similarly, statistical organizations in each country provide varying definitions of rurality.

In this paper we adopt another rurality measure. Inspired by the Swedish Agency for Growth Policy Analysis \cite{till2010}, we choose to create a rurality map based on the average minimum distance to the centers of cities or villages within specific population categories. This approach is particularly useful for Nordic countries, where population size and distribution significantly influence rurality assessments. In order to apply the CCI index, it is essential to measure rurality on a spectrum because we want to sort areas by their level of rurality rather than labeling them as simply rural or urban, suburban, or other grouping labels, similar to DEGURBA.

We start by considering a discrete set \(\mathcal{A}\) representing all locations in a region, think of this set as pixels of a map.  Each location \(x\) is defined by its coordinates, represented as longitude and latitude.  Then, we collect city center locations \(y\) of all cities with populations exceeding \(p\) in a set \(\mathcal{Y}^{(p)}=\{ y_1,y_2,\cdots ,y_{Y} \}\), where \(Y\) is the number of such cities, grouped by selected population sizes represented by the set $\mathcal{P}=\{ p_1,p_2,\cdots,p_N \}$, where \(N\) is the number of selected population sizes. The partial rurality of a location is then determined by its shortest distances to cities in these groups. The distances from any location \(x\) to the nearest city in each category are defined as
\begin{equation}
r^{(p)}(x)=  \min_{y\in \mathcal{Y}^{(p)}} d(x,y) \qquad \forall\ x \in \mathcal{A},
\label{e:partial_rurality_map}
\end{equation}
where \(d(x,y)\) is the distance measured for instance in kilometers. The equation, and $r^{(p)}$, represents what we call the \emph{partial} rurality map (distances to cities of population $p$). In \cite{till2010} urban areas in Sweden are classified by population sizes of at $\mathcal{P}=\{200, 1000, 3000,$\allowbreak $30000,60000\}$ \allowbreak inhabitants.

Using these partial rurality maps, we finally define the \emph{rurality map} for the region as the average of the partial rurality maps corresponding to all population sizes as
\begin{equation}
R(x)= \frac{1}{N}\sum_{p\in \mathcal{P}}r^{(p)}(x)  \qquad \forall x\in\mathcal{A}.
\label{e:rurality_map1}
\end{equation}
 
\begin{figure}[]
    \centering
    \includegraphics[width=\textwidth]{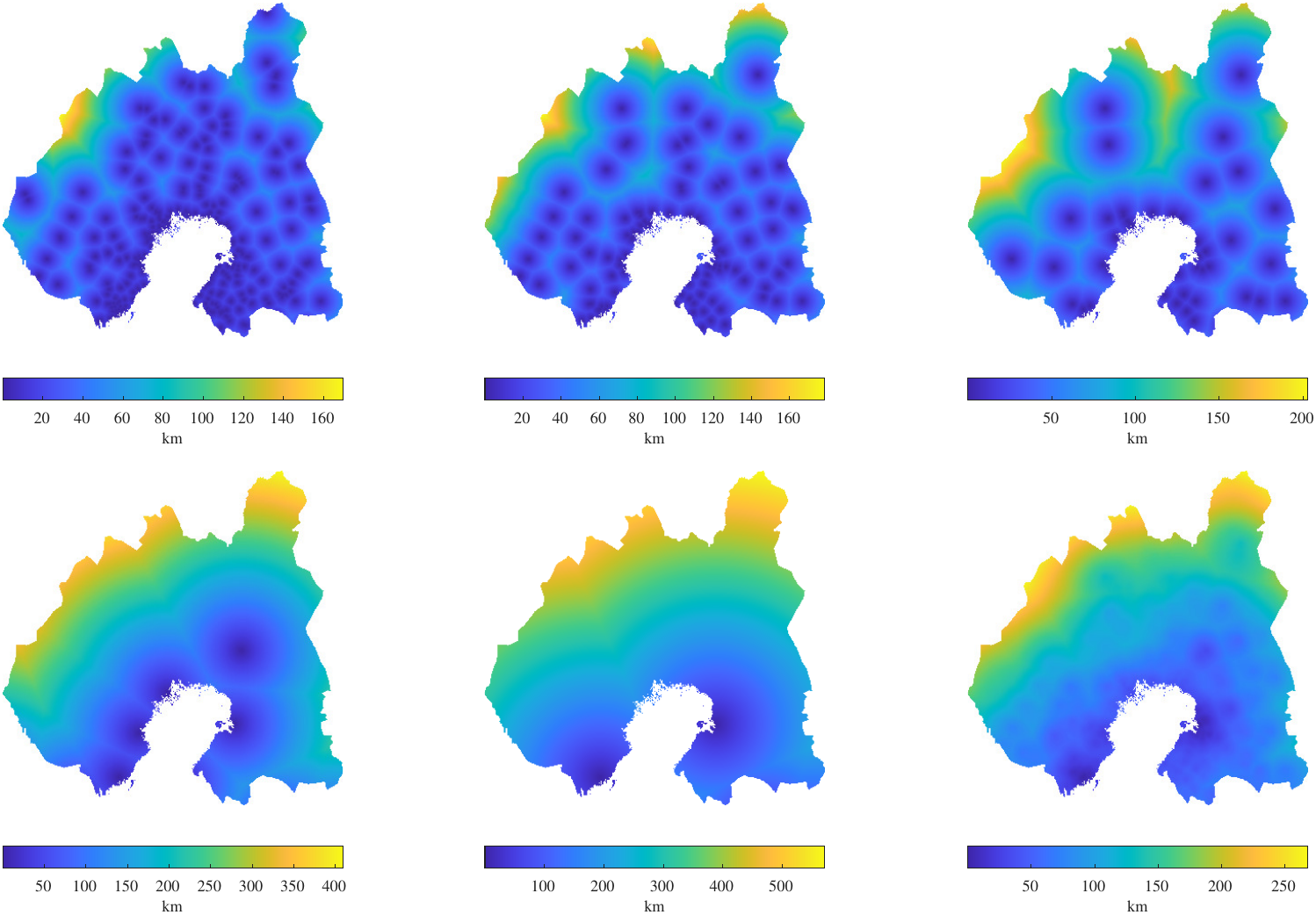}
    \caption{Five partial rurality maps and the resulting rurality map in the Arctic regions of Finland and Sweden for 2023. The partial rurality maps were generated by \eqref{e:partial_rurality_map} for a population size of 200 (upper left), 1000 (upper middle); 3000 (upper right), 30000 (bottom left) and 60000 (bottom middle). The resulting rurality map (bottom right) is created by \eqref{e:rurality_map1}.}
    \label{Rurality}
\end{figure}

Fortunately, the Nordic countries share a common definition of cities. The Swedish statistical agency, Statistiska centralbyrån (SCB), provides a list of urban centers in Sweden, including their populations and coordinates. A similar dataset is also available for Finland, provided by The Finnish Environment Institute, Suomen ympäristökeskus (SYKE). This shared definition and these lists allow the development of a joint rurality map for the Arctic regions across Sweden and Finland.  Such a joint rurality map significantly enhances our understanding of demographic trends, support regional planning efforts, and evaluate cellular coverage disparities.

We evaluate the joint rurality map of the Arctic region including Lapland as the northernmost region of Finland, along with North Ostrobothnia and Kainuu, as well as the northernmost region of Sweden, Norrbotten County and Västerbotten County.  For this joint Arctic region, Figure~\ref{Rurality} displays the five partial rurality maps $r^{(p)}$ for $\mathcal{P}=\{200, 1000, 3000, 30000,$ \allowbreak $60000\}$ \allowbreak alongside the final rurality map $R$, which represents the average of the five partial rurality maps. The partial rurality maps for 30000 and 60000 show a concentration of cities near the coastline of the Gulf of Bothnia. Consequently, the rurality map indicates that the least rural areas are located along the coastline, while rurality increases as we move further away from the coast and closer to the western and northern border with Norway.

\subsection{Cellular Coverage Map}\label{sec:4.2}
The second needed map shows the cellular coverage status of each location. Typically, mobile operators or regulatory agencies create and publish cellular coverage maps. These maps are usually binary, indicating whether coverage is available ('1') or not ('0'). They may be based on various factors, such as data rate, sensitivity (the ability of devices to detect signals), generation, and other relevant criteria.

Based on publications by the Finnish regulator, the Finnish coverage map is defined as
\begin{equation}
C_q(x) =
\begin{cases}
1 & \parbox[t]{9cm}{%
\text{if } \(x\) \text{ enjoys a data rate of (at least) 100 Mbit/s in the } \(q\) \text{-th generation},}\\[6pt]
0 & \text{otherwise,}
\end{cases}
\label{e:coverage_map_F}
\end{equation}

The Swedish regulator meanwhile publishes two slightly different coverage maps, defined as

\begin{equation}
C_b(x) = 
\begin{cases} 
1 & \text{if } x \text{ enjoys a data rate of at least } b \ \text{Mbit/s}, \\ 
0 & \text{otherwise,}
\end{cases}
\label{e:coverage_map}
\end{equation}

for $b=10, 30$ Mbits per seconds. Note the difference in coverage conditions across the two countries.

\begin{figure}[] 
\centerline{\includegraphics[width=\textwidth]{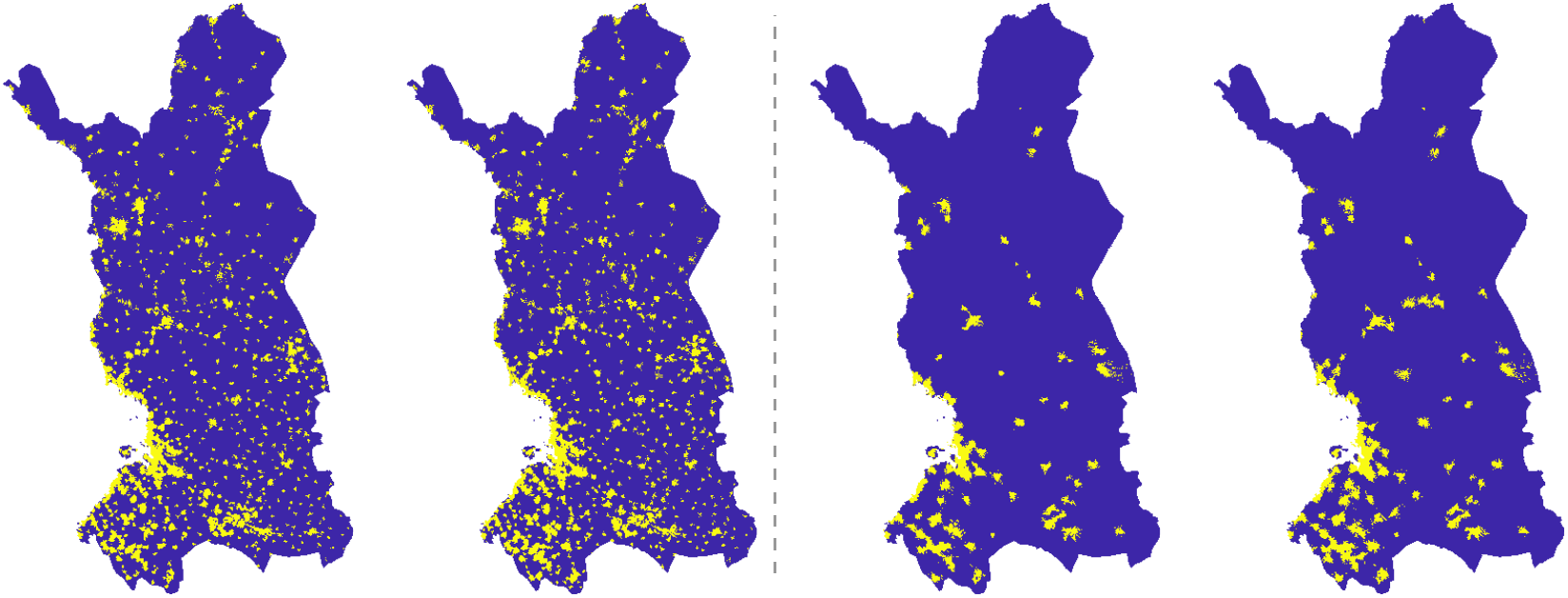}}
    \caption{Finland's 4G and 5G coverage maps created by  \eqref{e:coverage_map_F} for 2024: The left pair shows the 4G coverage map, displaying data rates of 100 Mbit/s and low sensitivity, with the far left representing 2024 and the middle left for 2025. The right pair shows the 5G coverage map, with data rates of 100 Mbit/s and low sensitivity, with the middle right for 2024 and the far right for 2025.}
    \label{Finland_Coverage}
\end{figure}

\begin{figure}[] 
\centerline{\includegraphics[width=0.33\textwidth]{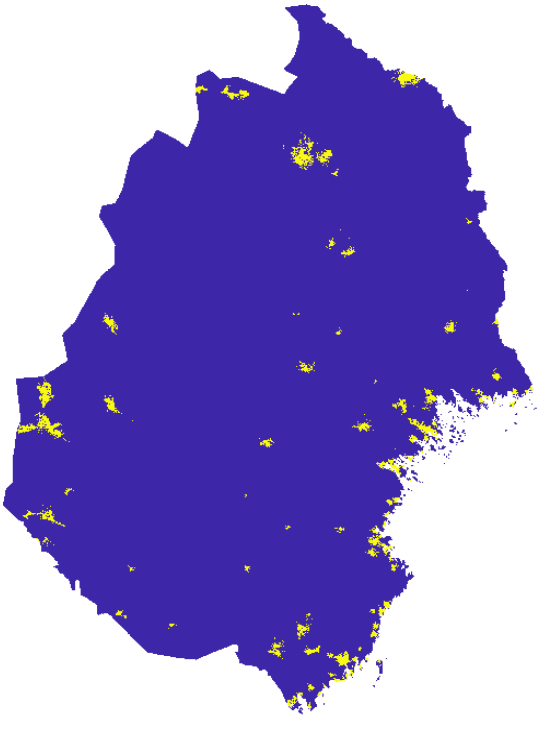}}
    \caption{Cellular Coverage Map of the Arctic Region of Sweden for a data rate of 30 Mbit/s and low sensitivity for 2019 created using \eqref{e:coverage_map}.}
    \label{Sweden_Coverage}
\end{figure}

Figure~\ref{Finland_Coverage} illustrates the cellular coverage map of Finland's Arctic regions as of September 2024. On the left, the map displays 4G coverage, characterized by data rates of 100 Mbit/s and low sensitivity. On the right, the map shows 5G coverage, which also features data rates of 100 Mbit/s and low sensitivity. This data is provided by the Finnish Transport and Communications Agency (TRAFICOM). Both maps have a resolution of 250 by 250 meters.  Figure~\ref{Sweden_Coverage} presents the cellular coverage map of Sweden’s Arctic region, highlighting areas with a data rate of at least 30 Mbit/s, characterized by low sensitivity for the year 2019. This data, sourced from the PTS, has a resolution of 250 by 250 meters.

These maps both illustrate that high data rate cellular coverage has primarily expanded in urban areas and their immediate surroundings. Furthermore, there are coverage around a few  holiday resorts located in the western part of the region.

\subsection{Cellular Coverage Inequality Index}\label{sec:4.3}

Based on the two maps (the rurality map and the coverage map), we outline a method to compute the CCI index, which was developed based on the work presented in \citep{CCI25}. The CCI index evaluates the distribution of cellular coverage in relation to rurality. The approach is inspired by concentration indices that measure how access to privileges in general varies among different socio-economic groups. However, instead of focusing on individuals, the CCI framework applies this concept to geographic areas, using rurality as the key variable.

The CCI index is one index value that reflects the rural-urban inequality for the whole region in a number range from 0 to 1. A value of 0 reflects perfect equality, meaning all areas have equal access to cellular coverage, while a value approaching 1 indicates significant disparities in coverage distribution  and a large urban-rural digital divide.

\begin{figure}[] \centerline{\includegraphics[width=.5\textwidth]{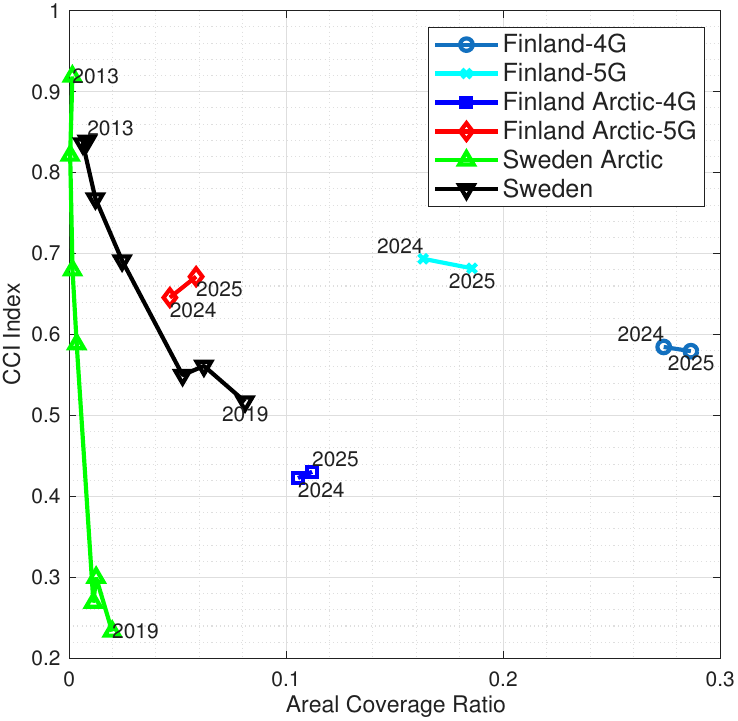}}
    \caption{Comparison of the Cellular Coverage Inequality (CCI) Index (y-axis) versus the Areal Coverage Ratio (ACR) (x-axis) for the national and Arctic regions of Sweden (2013–2019) and the national and Arctic regions of Finland (2024–2025).}
    \label{CCI_F_S}
\end{figure}

Figure~\ref{CCI_F_S} illustrates the CCI index against the traditional ACR for four distinct regions: Sweden (national), the Swedish Arctic, Finland (national), and the Finnish Arctic. For Sweden, the curve covering the 2013-2019 period, measured with low sensitivity and 30 Mbit/s data rates, shows that the CCI index followed a decreasing trajectory from 0.84 to 0.52, while the ACR grew from 0.84\% in 2013 to 8.11\% in 2019. In the Swedish Arctic, the ACR reflects a rise in coverage from 0.15\% to 1.99\%; however, this growth is not substantial, leaving the total coverage very low. Notably, the region's CCI index exhibited a sharp downward trend, dropping from 0.92 to 0.23 by 2019. The low ACR in Swedish Arctic makes it difficult to assess fairness trends, among other factors. 

In Finland, data from September 2024 to June 2025 for 4G and 5G networks with 100 Mbits/s data rates indicates that, for 4G networks, ACR increased from 27.39\% in 2024 to 28.65\% in 2025.
This slight rise in ACR was accompanied by a minimal decline in the CCI index, from 0.584 to 0.579, suggesting a marginal improvement in coverage equity. However, given the minimal change, this implies that the coverage distribution remained largely unchanged.

Similarly, for 5G networks, the ACR rose from 16.31\% to 18.56\% over the same period, while the CCI index decreased from 0.69 to 0.68. This also indicates a slight trend towards more equitable growth. Nonetheless, the changes are so small that they are effectively negligible, implying that coverage distribution fairness remained virtually the same.

Despite the slight national level expansion in coverage equity, the CCI index shows an opposite trend in the Arctic region. Although the ACR grew from 10.55\% to 11.20\% for 4G and from 4.64\% to 5.58\% for 5G between 2024 and 2025 are minimal and largely insignificant. this expansion disproportionately favored urban areas compared to 2024. This is reflected in the increase in the CCI index from 0.42 to 0.43 for 4G and from 0.65 to 0.67 for 5G. These increases signify a decline in coverage fairness relative to 2024. However, similar to the national level, the changes are so small that they remain practically negligible, suggesting that the overall fairness of coverage distribution in the Arctic region also remained almost unchanged.

It is important to note that the Finnish data covers only a nine-month period, which is insufficient to draw definitive conclusions or make reliable judgments about long-term trends. Additional historical data is needed to gain a more comprehensive understanding of coverage dynamics over time. Furthermore, the data periods and data rate standards differ significantly between Sweden and Finland, making it challenging to accurately compare the two neighboring countries regarding coverage fairness and trends. For example, Sweden's data spans a six-year period (2013-2019) with rates of 30 Mbit/s, while Finland's data covers only nine months (2024-2025) with rates of 100 Mbit/s for 4G and 5G. These inconsistencies highlight the need for harmonized data periods, standardized data rates, and consistent methodologies across regions to enable meaningful and fair comparisons of cellular coverage and its associated inequalities.

Additionally, there is an essential need for more publicly available data on cellular coverage to improve the accuracy and reliability of such analyses. Public datasets would ensure greater transparency and allow for better evaluations of trends in coverage distribution. Moreover, the development of standardized criteria for cellular coverage maps or the adoption of a common theoretical coverage model would greatly enhance the comparability of data between different countries. Without such standards, it becomes difficult to assess and compare coverage fairness effectively, even between neighboring countries like Sweden and Finland. These steps are crucial for accurately assessing rural-urban inequalities in cellular coverage and supporting more equitable digital access across diverse regions.

\section{Forward-Looking Perspectives on Connectivity}\label{sec:5}

Building on these shared challenges, the following section considers how residents and organizations in the Nordic Arctic envision the future of connectivity. Their perspectives highlight not only expectations for more stable and equitable services but also concerns about the pace and direction of digitalization. The results are presented in two parts: Section~\ref{sec:5.1} examines public expectations for future connectivity, while Section~\ref{sec:5.2} discusses visions and concerns related to emerging technologies in societal contexts.

\subsection{Public Expectations for Future Connectivity}\label{sec:5.1}

Survey responses revealed a consistent emphasis on the fundamentals of network provision: stability, coverage, and fairness of access. These were often discussed ahead of, and sometimes in opposition to, rapid technological rollouts. Thematic analysis uncovered four major expectations: reliability and stability first, closing coverage gaps for equity, fiber expansion for both access and backbone, balancing progress with affordability and security.

\theme {Reliability and Stability First}
Respondents consistently emphasized the importance of stable connectivity with improved reliability and data rates even re-iterating the existing connection failures despite multiple service providers operating in Finland. Alongside stability, respondents emphasized the need for faster connections. However, the preference was given for upgrading existing networks over investing in untested new technologies. As one participant argued \textit{“Would it be better to improve the existing ones that work than to try to create new ones with tens of millions, like 6G, which may or may not work.”}. One respondent even remarked that they had a better connection in Kenya than in Rovaniemi, underscoring regional inconsistencies in network reliability. Such frustrations highlight the need for regulatory targets and monitoring frameworks that go beyond nominal coverage maps and reflect real-world reliability benchmarks.

\theme{Bridging Coverage Gaps for Equity}
Many respondents emphasized the need to improve the current state of network coverage to be wider, ensuring reliability and quality. One of the key requirements was to have connections everywhere. As one participant put it: \textit{“Progress is good, but first the foundation and functioning connections must be created throughout the country.”} Some even pointed out that functioning networks covering all areas is more valuable than having data rates of 50 Mbit/s or even higher levels. Equal access to the internet regardless of location was emphasized especially at sparsely populated areas pointing out that people outside urban areas have the same needs so that connectivity should be developed in the same proportion in peripheral areas as in urban centers. Many expressed developments of at least 4G networks covering everywhere and some even preferring going back to 3G connections for wider coverage. While 3G services have now been discontinued in both Finland and Sweden, these responses reflect a broader concern that newer generations of mobile technology often fail to match the reliability and reach of earlier systems in rural contexts. These views highlight a public preference for closing persistent coverage gaps before pursuing the next generation of mobile technology, underlining the need for policy frameworks that prioritize spatial equity as much as technological advancement.

\theme{Fiber Expansion for Both Access and Backbone}
Respondents strongly emphasized the need for a wider and more accessible fiber network, calling for the expansion of fiber infrastructure to ensure reliable and high-capacity connectivity. They even reported of preferences of having fiber connections to every household and property so that the capacity demands do not have to be solely dependent on mobile networks. Responses made it much clearer that the public is skeptical about the capability of mobile networks to deliver higher data rates compared to fiber connections. Overall, the responses underscored that expanding fiber infrastructure is seen not as an optional enhancement but as a prerequisite for stable, high-speed connectivity and for equitable participation in digital society.

\theme{Balancing Progress with Affordability and Security}
Participants expressed mixed feelings about the rapid rate of digital advancements. Some even questioned whether continuous innovation is always beneficial. One respondent even expressed resentment towards the digital race expressing concerns about sustainability and the environmental impact. Some even expressed that they are satisfied with existing states while others prefer stability of connections over constant change. Another concern was affordability, highlighting the need for cost-effective solutions for cheaper and efficient connections. Interestingly bridging the skill gap of older adult populations were brought to attention highlighting the importance of inclusive and user-friendly digital services across all demographics. Finally, the need for stronger information security was a key concern. One participant expressed the need for information security developments at the same pace as internet connections to minimize network vulnerability due to cyber-attacks. \textit{“I hope that information security will develop at the same pace as internet connections, so that we will see fewer network crashes and other disruptions.”}. This reflects public awareness of cyber threats and the importance of secure and stable digital infrastructures. 

\medskip

Taken together, these perspectives highlight that connectivity strategies cannot be assessed only by data rate or coverage targets. Affordability, inclusivity, and security are equally critical for ensuring that digital infrastructures are trusted and sustainable.

\subsection{Future Technologies in Context}\label{sec:5.2}

Respondents envisioned a wide range of cutting-edge technological applications in future across professional, infrastructural and personal domains. Integration of artificial intelligence (AI) and immersive technologies were perceived to be shaping both professional and personal domains in freeing up routine tasks and bringing remote meeting experiences much closer to face-to-face encounters. Transport systems were expected to be automated and robotized with Mobility-as-a-Service (MaaS) model recuing requirements for private vehicle ownerships.  Respondents also pointed to advances in building automation and energy management, where greater use of sensors, remote control, and smart grids could improve efficiency and flexibility. In addition to applications, respondents reported a desire for multifunctional and durable devices designed for Arctic conditions. 

However, respondents also expressed reservations about the direction of technological change. One respondent expressed a desire for technology to be designed in ways that support user independence, wishing that apps and smartphones would not be created in ways that foster dependency. There was even a call for slowing down the pace of digital change and reconnect with life beyond the screen, with one participant stating: \textit{“I hope that digital development would already put on the brakes and people would start doing concrete work and being present in real life. So, I also hope that I myself will manage to take a step back and not fall for new applications so easily.”} The power of AI was not only recognized as a force shaping both professional and personal aspects of life but also as a potential source for social disruption. Another participant stated that they would not consider purchasing additional smart technologies until basic connectivity is secured, reflecting the importance of infrastructure over novelty. These concerns suggest that despite awareness of new opportunities, adoption will remain cautious unless reliable connectivity, balanced innovation, and trust in digital services are established.

\medskip

The forward-looking perspectives outlined in this section emphasize stability, equity, and trust in digital infrastructures, often giving these greater weight than rapid technological change. Opportunities in areas such as artificial intelligence, automation, and immersive applications are acknowledged, but respondents noted that such innovations will only become meaningful if basic connectivity gaps are first addressed. Together with the lived experiences reported in Section \ref{sec:3}, which highlighted frustrations with unreliable services, heightened safety risks, and feelings of exclusion in sparsely populated areas, these perspectives reveal a persistent misalignment between policy ambitions, technological developments, and everyday realities. Addressing this misalignment requires analytical tools and governance frameworks that move beyond conventional coverage metrics. The CCI index introduced in Section \ref{sec:4} provides one such analytical tool, complementing the qualitative findings by quantifying how rurality and network performance interact to shape unequal access to connectivity. Building on both the CCI index analysis and the experiential evidence presented above, Section \ref{sec:6} translates these insights into targeted policy reflections for advancing equitable broadband development.

\section{Policy Reflections}\label{sec:6}
The empirical findings, together with the CCI index analysis and the forward-looking perspectives presented in the preceding sections, show that the persistence of digital divides in northern Finland and Sweden arises from both structural and measurement shortcomings. Addressing these challenges calls for coordinated policy responses that combine regulatory flexibility with stronger transparency and accountability in monitoring. The following subsections outline key areas for intervention, drawing on lived experiences, quantitative evidence, and spatial indicators discussed earlier.

\subsection{Shared Infrastructure and National Roaming}\label{sec:6.1}
Findings from both the survey and field interviews (Sections \ref{sec:3.1}–\ref{sec:3.2}) show that connectivity problems in northern Finland and Sweden are not primarily due to a total absence of networks but to their limited reach and fragmented coverage across operators. Respondents described how mobile coverage could disappear within a few kilometers outside villages or along rural roads, and that reliable service often depended on which operator they used. In several areas, residents and entrepreneurs reported that one provider’s network functioned while another’s failed, forcing them to maintain multiple subscriptions. These accounts reveal that coverage gaps often exist between operators rather than in absolute terms, leaving users disconnected despite nearby infrastructure. This fragmentation demonstrates that isolated, operator-specific network development in low-density areas fails to ensure equitable and reliable coverage.

The underlying structural problem is that market-driven infrastructure expansion incentivizes duplication where population density is high but under-investment where it is low. Building parallel networks in remote Arctic terrain is technically possible but economically irrational, while regulatory frameworks in both Finland and Sweden largely assume operator-specific ownership models. 

A way to decrease infrastructure cost is infrastructure sharing, where multiple operators use the same towers or even the same radio network. This is a readily used approach. Another way to reduce cost is to share responsibility areas among operators requiring new legislation or regulation, since demands (in frequency licenses) for national operators must be changed. In this model an operator would be responsible for covering one area and another operator another area and so on. This solution also requires local national roaming so that people can connect in all shared areas. Local national roaming would also ease the operation of local operators that typically concentrate on improving coverage in a small area like a village. 

Comparable approaches already exist internationally. In the United Kingdom (UK), the \textit{Shared Rural Network (SRN)}, a government-operator programme overseen by Ofcom, aims to eliminate “partial not-spots” by encouraging infrastructure sharing \citep{Lavender2022,Ofcom2021}. In Australia, the \textit{Australian Competition and Consumer Commission (ACCC) } has examined “Temporary Disaster Roaming” as a contingency measure for maintaining mobile services during emergencies and rural capacity shortfalls \citep{ACCC2024}. While the ACCC concluded that such roaming is technically feasible, it also cautioned that mandatory roaming could reduce operators’ incentives for long-term investment. These findings highlight both the feasibility and the contested nature of sharing and roaming policies. This experience demonstrates that carefully designed frameworks, balancing cooperation and investment incentives, are essential for success.

\subsection{Spectrum Flexibility and Rural Coverage Models}\label{sec:6.2}

Local community-based operators might be interested in improving local coverage. An example in a Swedish mountain area shows this: In 2018, a first off-grid base station was installed near the top of a mountain near Lake Alesjaure, using a test-spectrum license and initially for non-commercial use only. The base station (which provided 2G and 4G connections) was placed in a container equipped with solar panels and batteries to provide power. Backhaul transmission was sourced from the main Mobile Network Operator (MNO) 4G network and connected to an own core network through a virtual private network (VPN) tunnel \citep{Imran2024}. Commercial spectrum in this region was initially difficult to obtain, as main national MNO's claimed exclusive rights through there expensively acquired national licenses anywhere in Sweden.  For decades, the MNO's licenses to use spectrum had not been used or exploited by the national operators (there has never been cellular coverage here) and the regulator later approved the use of the spectrum in a bounded region by a local network operator, illustrating that national spectrum licenses are not strictly and unlimitedly exclusive and that there are opportunities for local operators to start rural network businesses.

Finnish legislation allows local networks, though they are mainly targeted for industrial usage \cite{traficom_local-nets}. These are also limited to certain frequencies (2.3 GHz and 24.5 GHz in FI) that are not necessarily best suited for coverage extension. Since, in remote and rural areas, operators are not necessarily using all their allocated frequencies, these could be used by others, e.g., local operators. This may require changes in frequency licenses such that they become more flexible and allow utilization of unused spectrum. Alternatively, a licensing or agreement based approach may also be used. US based citizens broadband radio service (CBRS) \cite{FCCCBRS} is an example of this. Local networks already exist in Finland, but they currently operate only in limited frequency bands and are not used for coverage extension. So better suited frequencies made available through sharing, licensing or regulatory adjustments could further improve coverage.

Another complementary measure is the reverse-auction subsidy model, in which operators or consortia compete for the smallest public subsidy required to extend service into high-cost localities. This mechanism has been studied and applied in broadband-expansion programmes. For example, research on the United States \textit{Broadband Technology Opportunities Program (BTOP)} shows that, had reverse auctions been used instead of grant allocation, substantially more unserved households could have been connected within the same budget \citep{Oh2021}. The \textit{World Bank} \citep{WorldBank2023} has also examined the use of Multiple-Round Reverse Auctions (MRRAs), recommending their application to universal-service funds for rural broadband, with design work undertaken in Tanzania as a case example. In practice, the model has already been deployed at scale through the \textit{U.S. Federal Communications Commission Rural Digital Opportunity Fund Phase I (Auction 904)}, where competitive bidding determined the lowest subsidy required to extend broadband in underserved areas \citep{FCC2020}. The same approach could be adopted in Arctic municipalities where deployment costs are prohibitive.

\subsection{Quality of Service as a Measure}\label{sec:6.3}

Empirical findings from Sections \ref{sec:3.1}–\ref{sec:3.2} show that the quality of mobile connectivity in rural and Arctic Finland and Sweden is constrained less by complete absence than by irregular and unreliable performance across time and place. Respondents described wide variations in speed and stability depending on location and network load. Even within small settlements, service could be strong at one point and unusable a few hundred meters away. Several interviewees reported that connections slowed or failed during evening peaks or tourist seasons and that calls or data sessions frequently dropped when moving between coverage zones. 

Current monitoring and regulatory frameworks in Finland and Sweden rely primarily on nominal coverage and data-rate thresholds. These headline figures overstate inclusiveness because they measure potential access rather than actual performance. The absence of time-of-day or location-specific indicators means that temporary congestion and reliability shortfalls go undetected, while official statistics continue to show full coverage. In many rural and tourism-dependent areas, network capacity during peak periods is not sufficient to maintain stable service, yet such shortfalls remain invisible in current evaluation systems. Consequently, rural and Arctic users remain unrepresented in national monitoring despite facing the most severe service degradation.

To ensure that connectivity translates into meaningful participation in digital life, broadband policy must shift from static availability measures toward performance-based evaluation. Quality of Service (QoS) should be treated as a core policy metric, not only a technical indicator, because it captures the usability and reliability of digital access. This requires new KPIs that reflect how networks perform under real conditions, including temporal fluctuations and capacity constraints.

Four candidate indicators illustrate such an approach. \textit{Location reliability} measures whether even the least well-served points within a municipality meet a minimum standard for data rate and latency, highlighting local consistency rather than averages. \textit{Time-of-day capacity} addresses peak-hour congestion by comparing typical and lower-end performance, particularly during evenings and seasonal surges. \textit{Session continuity} reflects the ability to sustain uninterrupted connectivity for tasks such as remote work or distance learning, an issue repeatedly reported in the field. Finally, \textit{safety-critical access} measures the reliability of emergency communications under high-load conditions, establishing a benchmark for dependable service in sparsely populated areas.

While such indicators are conceptually straightforward, their practical implementation is constrained by data quality. Current coverage maps in Finland and Sweden are largely based on simulation models rather than systematic field measurements, which raises questions about their reliability and accuracy. As a result, existing data sources such as operator reports and regulatory maps do not provide an adequate empirical foundation for detailed performance indicators. Developing meaningful and credible measures therefore requires more comprehensive and transparent data collection, including independent measurements and verification. Once such data become available, integrating spatial inequality metrics such as the CCI index would enable regulators to detect and correct systematic disparities between urban and rural regions more effectively.

International policy analysis reinforces this view. If the goal of broadband policy is to ensure equitable participation in education, health, commerce, and public services, then performance metrics must be redefined to reflect how people actually depend on connectivity in everyday life \citep{OECD2025,BEREC2021}. The Body of European Regulators for Electronic Communications (BEREC) has similarly emphasised that national regulators should complement operator-reported maps with independent quality-of-service measurements, including speed, latency, and reliability at user locations. Aligning Nordic frameworks with these principles would enhance both accuracy and accountability.

Not only QoS is needed as a new measure of connectivity, but connectivity figures based on it should be reported clearly. Furthermore, connectivity maps (based on QoS) of different operators should be easily comparable to ease selection process. Today, in practice, coverage reporting across Finland, Sweden, and the wider EU remains fragmented. National statistics use different thresholds, map resolutions, and definitions of coverage, making direct comparison difficult. Furthermore, in many cases data is not openly available. Even when data are available, the accompanying metadata are often incomplete or inconsistent, leaving unclear how measurements were made or when they were last updated. Without transparent documentation of measurement methods, it becomes impossible to interpret figures or link them to real-world performance. Improving the interpretability and comparability of coverage data is therefore essential for credible broadband policy. Standardisation should not only concern the indicators themselves but also the methodological context that allows users, municipalities, and researchers to understand them. Metadata must describe data sources, validation procedures, and measurement methods so that reported coverage can be meaningfully evaluated.

\subsection{Digital Skills and User Practices}\label{sec:6.4} 

One approach to ease connectivity problems is to show how existing connections can be used more efficiently. Limited connectivity is not always the main constraint; it can also reflect the lack of clear, practical information on how to improve or assess existing connections. While some people learn these skills through experience or shared advice online, not everyone has this knowhow, and the quality of available guidance varies widely. Authorities therefore have an important role in maintaining up-to-date, reliable, and easily accessible information on how to make the most of existing infrastructure. This guidance should be tested rather than promotional, practical rather than generic, and visible through official communication channels. Examples include instructions on:

\begin{itemize}
\item Using different gadgets to improve connectivity, such as modems or mobile modems with and without external antennas.
\item Locating and interpreting operators’ coverage maps to check connectivity in specific areas and to compare operators’ performance.
\item Understanding how to submit complaints to authorities when mandatory service quality standards are not met, with the relevant quality metrics shown in the same place.
\item Using offline features of applications and services, such as maps, in areas with poor or unstable coverage.
\end{itemize}

Maintaining this information base and testing solutions requires sustained effort and coordination. Yet such initiatives are often constrained by limited public sector resources. Incorporating structured information services into broadband and digital inclusion programmes would help ensure that connectivity improvements are supported not only by infrastructure but also by informed and capable users.

\section{Conclusions}\label{sec:7}

This study examined digital connectivity in the northern regions of Finland and Sweden by combining survey and interview evidence with a spatial analysis of mobile coverage using the CCI index. The mixed-method approach captured both lived experiences and measurable disparities and linked these findings to policy reflections on broadband governance. The results show that despite high national levels of digitalization, rural and Arctic communities continue to face uneven and unreliable connectivity. Everyday activities, whether personal communication, business operations, or safety-critical tasks, remain constrained by gaps in coverage and by variation in network performance across time and place. These disparities are not only technical but institutional, reflecting how broadband targets, monitoring practices, and market structures shape the quality and distribution of digital access.

The analysis demonstrates that bridging the urban–rural digital divide requires a policy approach that extends beyond infrastructure provision alone. Fibre, cellular, and satellite networks each serve complementary purposes. Stable digital access depends on combining fixed capacity with well-functioning mobile systems. Respondents viewed fibre as the most dependable foundation, particularly when mobile networks become congested. This points to the need for continued fibre expansion, both for household access and as backbone support for wireless services. 

At the same time, connectivity in sparsely populated areas cannot rely solely on market incentives. Shared infrastructure, responsibility area sharing, and local national roaming can reduce costs and extend coverage where parallel networks are not economically feasible. Allowing micro-operators, including local or municipal actors, to participate through flexible spectrum allocation could further improve rural coverage. This will require regulatory frameworks that enable access to unused frequencies and closer coordination between operators and regulators. Together, these measures align with broader European goals for inclusive and efficient network deployment.

Measurement practices also need revision. Headline figures such as covered population percentage (e.g., although saying that 99\% of a 5 million population are covered by 4G sounds great, it means that 50,000 people are unconnected) and coverage area percentage often overstate inclusiveness and mask temporal congestion and location-specific weaknesses. The CCI index introduced in this study provides a complementary tool for assessing how coverage expands across degrees of rurality and for identifying areas where connectivity remains least equitable. Incorporating QoS measures would offer a more realistic assessment of network performance. We identified four candidate indicators for this purpose: location reliability, time-of-day capacity, session continuity, and safety-critical access. Harmonized and openly available coverage data would allow comparison across operators, regions, and countries, and enable independent assessment by researchers, regulators, and policymakers. Transparency in measurement is therefore a precondition for accountability in broadband policy.

Finally, the findings underline that digital inclusion depends not only on infrastructure but also on users’ ability to make effective use of existing connections. Limited connectivity is not always the main constraint; it can also result from the lack of clear and practical information on how to assess or improve network performance. While some people learn these skills through experience or advice shared online, many lack reliable guidance, and the quality of available information varies widely. Authorities therefore play an important role in maintaining reliable and easily accessible guidance on how to use existing infrastructure efficiently. This information should be practical, tested, and visible through official channels rather than promotional or generic. Incorporating such structured information services into broadband and digital inclusion programmes would help ensure that connectivity improvements are supported not only by infrastructure but also by informed and capable users.

Taken together, these results and policy reflections suggest that equitable digital access in the Nordic Arctic requires an integrated approach that combines reliable and diverse infrastructure, shared and flexible governance of networks and spectrum, quality-focused monitoring, and informed users. Viewing connectivity as a dynamic public good whose quality, availability, and usability must be jointly maintained provides a practical foundation for reducing spatial digital inequalities in Finland, Sweden, and other sparsely populated regions.

\section*{Acknowledgement}
This work was partially funded by the InterReg Aurora program through the "Arctic 6G" project. It was financially supported by Interreg Aurora (EU), Lapin Liitto (FI) and Region Norbotten (SE). It was partially funded by the Academy of Finland 6G Flagship (grant 318927). The first author was fully funded by the Infotech Emerging Project under the University of Oulu.

\section*{Declaration of Generative AI and AI-Assisted Technologies in the Writing Process}
During the preparation of this manuscript, the first author used ChatGPT to correct grammar, refine language, and adjust the academic tone. The first author reviewed and edited all AI-generated content and takes full responsibility for the final version of the manuscript.

\bibliographystyle{unsrtnat}
\bibliography{main}

\end{document}